\documentclass[12pt]{article}
\usepackage[english]{babel}
\usepackage{amssymb,latexsym,epsfig,cite}
\usepackage{graphicx}
\usepackage{fix-cm}

\newcommand{\be}{\begin{equation}}
\newcommand{\ee}{\end{equation}}
\def\bea{\begin{eqnarray}}
\def\eea{\end{eqnarray}}
\begin{document}

%%%%%%%%%%%%%%%%%%%%%%%%%%%%%%%%%%%%%%%%%%%%%%%%%%

\begin{center}
\Large{\bf Negative time delay for wave reflection from a one-dimensional
semi-harmonic well}
\end{center}

\vskip2ex
\begin{center}
Oscar Rosas-Ortiz${}^1$, Sara Cruz~y~Cruz${}^2$ and Nicol\'as Fern\'andez-Garc\'{\i}a${}^2$\\[2ex]
${}^1$ Departamento de F\'{\i}sica, Cinvestav,\\
AP 14-740, 07000 M\'exico~DF, Mexico\\[1ex]
${}^2$ {\em Secci\'on de Estudios de Posgrado e Investigaci\'on, UPIITA-IPN,\\
Av. IPN 2580, CP 07340 M\'exico~DF, Mexico}
\end{center}

\vskip2ex
\begin{center}
\begin{minipage}{14cm}
\begin{center}
{\em \footnotesize To Professor Bogdan Mielnik with our deepest admiration.}
\end{center}

\vskip1ex
\noindent
{\bf Abstract} It is reported that the phase time of particles which are
reflected by a one-dimensional semi-harmonic well includes a time
delay term which is negative for definite intervals of the
incoming energy. In this interval, the absolute value of the
negative time delay becomes larger as the incident energy becomes
smaller. The model is a rectangular well with zero potential
energy at its right and a harmonic-like interaction at its left.
\end{minipage}
\end{center}

%%% ----------------------------------------------------------------------
%\maketitle
%%% ----------------------------------------------------------------------
%\tableofcontents

%%% ----------------------------------------------------------------------
\section*{}
\vskip-2ex

The time taken by a particle to traverse a given spatial region is
one of the most striking features of quantum theory
\cite{Mug08,Mug10}. In the case of tunneling through a
one-dimensional barrier of height $V_0$ and width $\xi$, the
\textit{transmission time} of a wave packet centered at
the average total energy $E= \hbar \omega= \hbar^2 k^2/(2m) <V_0$
is independent of the barrier thickness \cite{Har62}. Thus, the
peak value of the packet propagates with the effective group
velocity $v_g=d\omega/dk=\hbar k/m$, which must increase with
$\xi$ across the barrier. Using electromagnetic analogues,
superluminal (``anomalously large'') group velocities have been
observed for evanescent modes \cite{Mar92}, microwave pulses
\cite{End92}, and in the tunneling of photons through 1D photonic
band gaps \cite{Ste93}. Indeed, this `abnormal behavior' of light
\cite{Bri60} has stimulated the designing of high-speed devices
based on the tunneling properties of semiconductors (see, e.g.,
Chs. 11 and 12 of Ref. \cite{Mug08}). In the stationary phase
approximation \cite{Wig55}, the \textit{phase time} (group delay)
is defined as the energy derivative of the transmission phase
$\tau_W= \hbar \frac{d \varphi}{dE} = \frac{1}{v_g} \frac{d
\varphi}{dk}$. This gives information of the time taken by the
peak of the transmitted packet to appear, measured from the moment
the peak of the incident packet strikes a given barrier. Another
well established notion of time considers the average time spent
by the particles in the barrier. It is called the \textit{dwell
time} and is defined as the ratio $\tau_D = n/j$, with $n$ the
number of particles within the barrier and $j$ the incident flux
\cite{Smi60}. Yet, $\tau_W$ and $\tau_D$ are not necessarily
related with each other; they are comparable only if the barrier
is almost transparent \cite{But83}.

While the quantum tunneling of rectangular barriers has attracted
a lot of attention in recent years (see e.g., \cite{Hua89,Lan94}
and references quoted therein), the scattering properties of
rectangular wells have been underestimated. Quite recently,
however, nonevanescent propagation has been predicted for
potential wells \cite{Li00}. In contradistinction with the
tunneling exponential attenuation, the scattering at quantum wells
attenuates the outgoing wave packets only because of the multiple
reflections at the well boundaries. Negative phase times are then
expected under certain conditions of the incident energy and the
thickness of the well \cite{Li00,Alo11}, a phenomenon which should
be observable for electromagnetic wave propagation \cite{Vet01}.
Thereby, rectangular wells may lead to much larger advancements
than rectangular barriers in the context of traversal times
\cite{Mug02}.

%%%%%%%%%%%%%%%%%%%%%%%%%%%%%%
\vskip3ex
\begin{figure}[htb]
\centering\includegraphics[width=11cm]{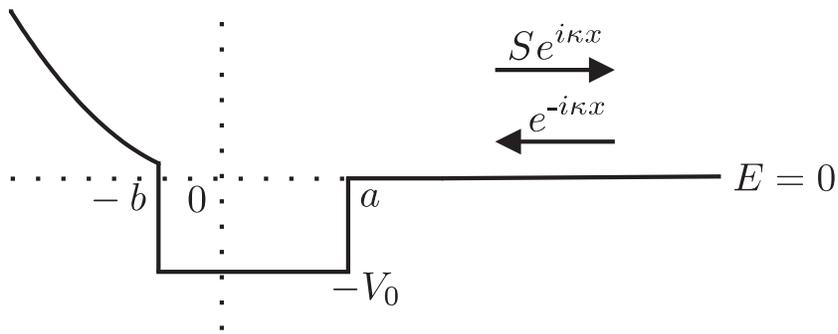}

\caption{\footnotesize Schematic representation of the one-dimensional
semi-harmonic square well as a function of the dimensionless
position $x$. The wave $e^{-ikx}$ colliding the well from the
right is reflected to give $Se^{ikx}$, with $S=e^{i\delta}$ the
reflection amplitude and $\delta(E)$ the reflection phase shift.}
\label{potential}
\end{figure}
%%%%%%%%%%%%%%%%%%%%%%%%%%%%%%

The purpose of this contribution is to report negative time delay
for a one-dimensional well which reduces the scattering process to
the case of purely wave reflection. The absolute value of this
negative time delay becomes larger as the energy of the incident
particle becomes smaller. To begin with, consider the stationary
Schr\"odinger equation $(H-E) \psi(x)=0$, where $V(x)$ is the
one-dimensional potential depicted in Figure~\ref{potential}. This
last is a rectangular well in a semi-harmonic background
integrated by zero potential energy (flat potential) at the right
and a harmonic-like potential at the left of the well. Our model
corresponds to a system (the rectangular well) embedded in an
environment (the parabolic plus flat potentials), and the issue is
the study of the modifications on the physical properties of the
system due to the environment \cite{Fer11}. For instance, the
number $N+1$ of bound states $\psi_n(x)$, $n=0,1,\ldots,N$, is
determined by the area $A = (a+b)V_0$ of the rectangular well.
Here, $a+b$ and $-V_0$ are respectively the width and depth of the
well with $V_0>0, a \geq 0$, and $b \geq 0$. Once the
semi-harmonic background is added, the number $N+1$ is preserved
but the corresponding energies $E_0, E_1, \ldots, E_N$, are
displaced towards the positive threshold. This last property does
not depend on the geometry of the rectangle; the wells having the
same area admit the same number of bound states. In this context,
remark that the wells of unit area $V_0=a+b$ admit only one bound
state and constitute a family of compact support functions which
converge to the delta well in the sense of distribution theory
\cite{Neg02}. Then, the single bound state (dimensionless) energy
$E_0=- 0.25$ of the delta well becomes less negative $E_0 = -
0.0797104$ in the presence of the semi-harmonic background
\cite{Fer11} (compare with \cite{Esp08}).

%%%%%%%%%%%%%%%%%%%%%%%%%%%%%%
\begin{figure}[htb]
\centering\includegraphics[width=5cm]{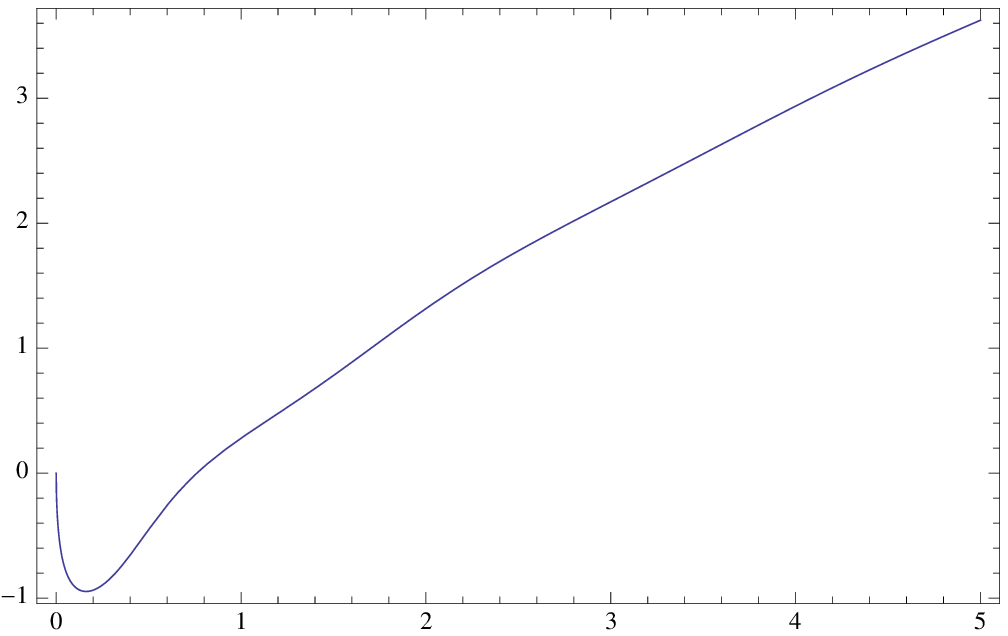} \hskip1cm
\includegraphics[width=5cm]{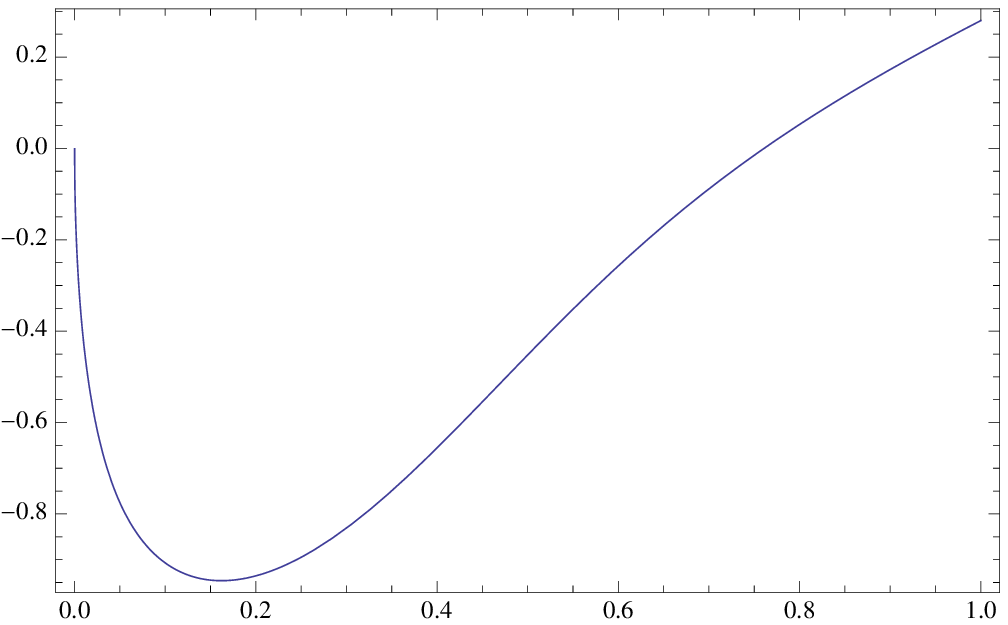}

\caption{\footnotesize The reflection phase shift $\delta(E)$ of a semi-harmonic
well of unit area as a function of the dimensionless energy $E$
for the parameters $a=b=5/2$, and $V_0=1/5$. A detail of the
behavior of $\delta(E)$ for low energies is shown at the right.}
\label{phase}
\end{figure}
%%%%%%%%%%%%%%%%%%%%%%%%%%%%%%

On the other hand, the isolated resonances of a rectangular well
are easily identified by expressing the transmission amplitude $T$
as a superposition of Breit-Wigner distributions \cite{Fer08}. The
center $E^r_k>0$ and width $\Gamma_k$ of each of these peaks
define the resonance complex eigenvalue $\epsilon_k = E^r_k
-i\Gamma_k/2$, and induce time delays in the scattering process
\cite{Ros08}. A rapid increasing of the transmission phase is then
expected in the vicinity of the resonance position $E_k^r$.
According to Wigner, the increases of the phase should be balanced
by the appropriate decreases \cite{Wig55}. Therefore, the slope of
the transmission phase can be even negative in order to compensate
for the phase increases associated with each of the resonances.
This effect is more important near the energy threshold, below the
position of the first Breit-Wigner peak of $T$ \cite{Alo11}. In
other words, the negative phase times predicted in \cite{Li00} are
in complete agreement with the conditions to get at least one
isolated resonance in rectangular wells \cite{Alo11,Fer08}. If the
semi-harmonic environment is activated, all the scattering states
become more excited and their wave functions cancel at
$x=-\infty$. As the potential includes neither sources nor
shrinks, the probability is conserved and all the incoming waves
are reflected. Then, the reflection phase shift $\delta(E)$
encodes all the information of the scattering process. This phase
is depicted in Fig.~\ref{phase} for a unit area semi-harmonic
square well with $a=b=5/2$. Notice the strong negative slope in
the interval of dimensionless energies $(0, 0.16208517)$, so that
negative time delay is expected for wave packets colliding the
well from the right at the appropriate energy.

The straightforward calculation shows that the phase time is given
by $\tau_W= \tau_p -\tau_E$, with $\tau_p = 2a/v_g$ the classical
flight time to traverse a distance $2a$, and the time delay
$\tau_E$ written in the form
\[
\tau_E= \frac{1}{v_g} \frac{\partial}{\partial k} \left[ \arctan
\left( \frac{2 \phi_1 \phi_2}{\phi_1^2 - \phi_2^2} \right)\right].
\]
Here the functions $\phi_1$ and $\phi_2$ are given by
\[
\phi_1 = -\frac{q}{2} \sin 2qa + \left. \frac{\psi'}{\psi}
\right\vert_{x=-a} \cos 2qa, \qquad \phi_2= -k\cos 2qa -\left.
\frac{1}{q} \frac{\psi'}{\psi}\right\vert_{x=-a} \sin 2qa,
\]
with
\[
\psi(x)= e^{-x^2/2} \left[{}_1F_1 \left(\frac{1-k^2}{4}, \frac12;
x^2 \right) +2x
\frac{\Gamma(\frac{3-k^2}{4})}{\Gamma(\frac{1-k^2}{4})} {}_1F_1
\left(\frac{3-k^2}{4}, \frac{3}{2}; x^2 \right)\right],
\]
and $q=\sqrt{V_0+k^2}$. The expression ${}_1F_1(a,c;z)$ stands for
the confluent hypergeometric function.

%%%%%%%%%%%%%%%%%%%%%%%%%%%%%%
\vskip3ex
\begin{figure}[htb]
\centering\includegraphics[width=7cm]{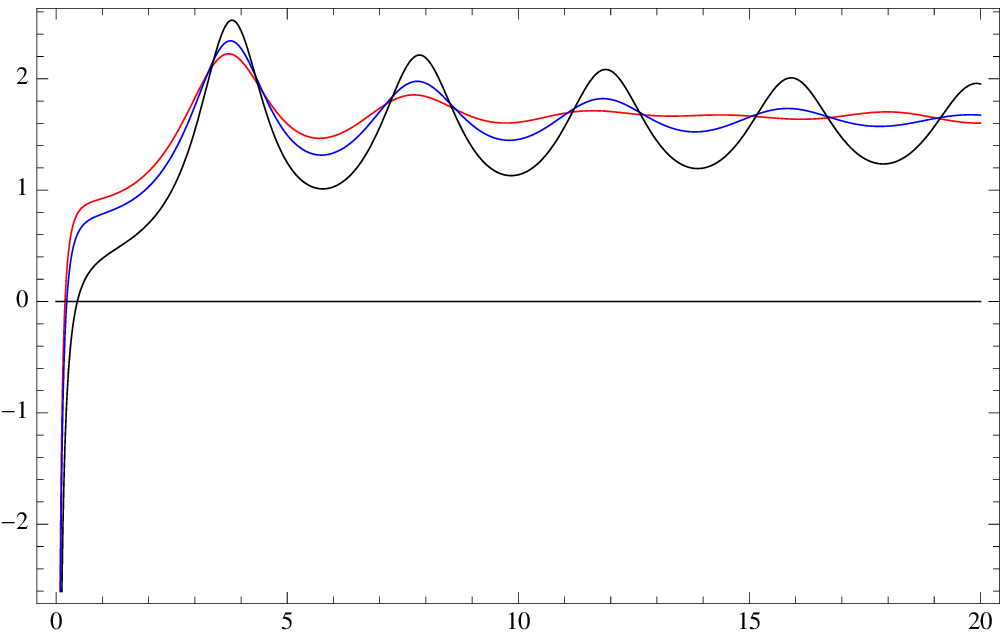} \hskip1cm
\includegraphics[width=7cm]{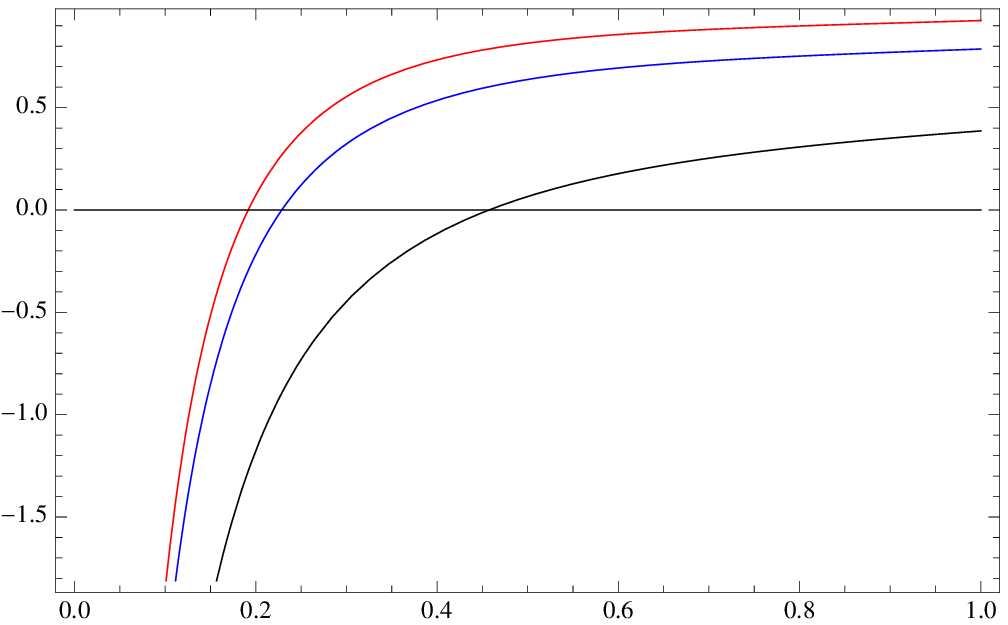}

\caption{\footnotesize Time delay $\tau_E$ of a unit area semi-harmonic well for
$a=0.4$ (red curve), $a=0.03$ (blue curve) and $a=0$ (black
curve). A detail of the behavior at low energies is shown at the
right.}
\label{delay}
\end{figure}
%%%%%%%%%%%%%%%%%%%%%%%%%%%%%%

%%%%%%%%%%%%%%%%%%%%%%%%%%%%%%%%%%%%%%%%%%%%%%
\begin{table}
\begin{center}
\begin{tabular}{|c|c||c|c|}
\hline
$a$ & $E_a$ & $a$ & $E_a$\\
\hline\hline
2.5 & 0.03406092 & 1.0 & 0.10100123\\
2.0 & 0.05056413 & 0.5 & 0.16473112\\
1.5 & 0.07205970 & 0.0 & 0.45727096\\
\hline
\end{tabular}
\end{center}
\caption{\footnotesize The (dimensionless) energy $E_a$ defining the change of
sign in the time delay for a semi-harmonic well of unit area.}
\label{table}
\end{table}

%%%%%%%%%%%%%%%%%%%%%%%%%%%%%%%%%%%

Figure~\ref{delay} shows the behavior of the time delay $\tau_E$
for some semi-harmonic wells of unit area but different
geometries. Given $a$, there is an interval of scattering energies
$(0,E_a)$ where $\tau_E$ is negative (for definite values see
Table~\ref{table}). In this interval, the absolute value of the
negative time delay becomes larger as the incident energy becomes
smaller. Thus, it is clear the dependence of $\tau_E$ on the
energy $E$ of the incident particles and on the rectangular well
thickness $2a$. For a given value of $a$, the maxima of the time
delay are localized at the real part of the resonance eigenvalues
$\epsilon_k =E^r_k-i \Gamma_k/2$, as expected. The energies
$E_k^r$ are displaced to more excited values as $a \rightarrow 0$.
In the very limit $a=0$, the time delay changes its sign at the
scattering energy $E_{a=0} = 0.45727096$ and oscillates around the
asymptotic value $\frac{\pi}{2}$ for $E>E_{a=0}$. It should be
pointed out that the interval of scattering energies $(0,E_{a=0})$
is the largest one in which $\tau_E$ is negative for any of the
unit area semi-harmonic wells (see Table~\ref{table} and
Figure~\ref{delay}).

Let us close this contribution with some remarks on the optical
analogs applied in the study of particles passing through a
rectangular well \cite{Li00,Vet01}. Of particular interest,
negative phase times have been confirmed for electromagnetic wave
propagation in waveguides filled with different dielectrics
\cite{Vet01}. The negative time delay $\tau_E$ of the
semi-harmonic wells could be studied in a similar way by taking
$b=0$ and $a \geq 0$. Once the energy baseline of the rectangular
well is shifted by the constant value $E_0=\hbar \omega_0$, the
cutoff frequency $\omega_0$ of the first waveguide section is
defined. Then, waveguide sections with different cutoff
frequencies can be constructed to approximate the parabolic part
of the potential by a series of Riemann rectangles. As a result,
the semi-harmonic well can be connected to a piecewise frequency
$\omega_c(x)$. Following \cite{Vet01}, the solution to the
propagation problem (i.e., the Helmholtz equation for $\omega_c$)
is obtained if the wave functions and the electromagnetic fields
satisfy identical boundary conditions. Further details will be
given elsewhere.

% ------------------------------------------------------------------------
\vskip12pt
%\subsection*{Acknowledgment}
This research was supported by CONACyT under grant 152574, and by
the IPN grants SIP20113705 and SIP20111061. ORO wishes to thank the 
Organizers of the Conference ``XXX Workshop on Geometric Methods in Physics''
for the kind invitation to give a talk in the
\textit{Special Session in honour of Bogdan Mielnik}, and for the
warm hospitality at Bia\l owie\.{z}a Forest.

% ------------------------------------------------------------------------
\end{document}